\documentclass[aps,prl,twocolumn,groupedaddress]{revtex4-1}

\usepackage{graphicx}
\usepackage{bm}
\usepackage{color}
\begin{document}


\title{Helical locomotion in granular media}


\author{Baptiste Darbois Texier}
\author{Alejandro Ibarra}
\author{Francisco Melo}
\affiliation{SMAT-C and Departamento de F\'isica Universidad de Santiago de Chile, Avenida Ecuador 3493, 9170124 Estaci\'on Central, Santiago, Chile}


\date{\today}

\begin{abstract}

The physical mechanisms that bring about the propulsion of a rotating helix in a granular medium are considered. A propulsive motion along the axis of the rotating helix is induced by both symmetry breaking due to the helical shape, and the anisotropic frictional forces undergone by all segments of the helix in the medium. Helix dynamics is studied as a function of helix rotation speed and its geometrical parameters. The effect of the granular pressure and the applied external load were also investigated. A theoretical model is developed based on the anisotropic frictional force experienced by a slender body moving in a granular material, to account for the translation speed of the helix. A good agreement with experimental data is obtained, which allows for predicting the
helix design to propel optimally within granular media. These results pave the way for the development of an efficient sand-robot operating according to this mode of locomotion.
\end{abstract}


\maketitle


Achieving locomotion within a fluid at low Reynolds number requires that the mobile parts of swimmers mobilize in a non-reciprocal motion, as stated by the Purcell's scallop theorem \cite{purcell1977life}. An efficient way of achieving propulsion is through the rotation of a helical filament \cite{berg1973bacteria}, equivalent to a propagating wave, which breaks the inversion symmetry of the system. In nature, the helical strategy is widely adopted by microorganisms such as \textit{E. Coli} and \textit{C. Crescentus} \cite{lauga2009hydrodynamics}. Due to its fundamental importance, the helical motion of \textit{E. Coli} has been investigated extensively. Specifically, by using optical traps, it has been possible to assess the forces and torques generated by \textit{E. Coli} in vivo \cite{chattopadhyay2006swimming}. From a theoretical point of view, helical propulsion in viscous fluids was first addressed by Lighthill using the resistive-force theory \cite{lighthill1976flagellar}. This approximation accurately predicts the propulsive mechanisms, however it neglects the effect of the long-range hydrodynamic interactions between different parts of the body. Recently, applying slender body theories and the regularized Stokeslet method, Rodenborn \textit{et al.} have tested these approximations \cite{rodenborn2013propulsion}.

 Under a technological context, helical propulsion has acted as a natural source of inspiration for the design of self-propelled micro-robots, for example, set in motion through external magnetic fields \cite{zhang2009artificial}. However, the production of helical micro-tails require creative strategies; for instance, the bending of a thin bilayered film with pre-constraints, and the direct 3D fabrication of components through laser writing into a photoresist \cite{nelson2010microrobots}. Despite challenging technological issues, micro-robots have facilitated the development of various biomedical applications, ranging from minimally invasive surgery, to targeted drug delivery and cell manipulation \cite{peyer2013magnetic}. Helical motion has also shown to be efficient for propulsion in non-newtonian fluids \cite{liu2011force,spagnolie2013locomotion,gomez2017helical} and drop climbing \cite{texier2015droplets}.

\noindent Furthermore, a recent study demonstrates that helical rotation is nicely employed by \textit{Erodium} and \textit{Pelargonium} seeds to penetrate cohesionless soils \cite{jung2017reduction}. Besides, the use of the helical mode of locomotion among the animal kingdom at a macroscopic scale remains an open-ended question.

Purcell's scallop theorem has not been proven within granular materials. However, on an entirely different length scale, many animals develop propulsion strategies similar to that of viscous fluids. For example, sand lizards and snakes are able to propel up to twice their body length per second both under and over a sandy surface. This is achieved through the propagation of a transverse wave \cite{maladen2009undulatory}. Moreover, despite differences in the physical mechanisms involved, a solid friction analog to the resistive-force theory in viscous fluid has been successfully applied in the context of granular media to describe the undulatory motion of sand lizards and snakes \cite{maladen2011mechanical,hu2009mechanics}. In addition, Malden \textit{et al.} demonstrated that this is due to the anisotropy of the friction force, produced by a slender body moving in a granular medium \cite{maladen2009undulatory,ding2012mechanics}; the frictional forces per unit of length are greater for motion occurring perpendicular to the body compared to along the body.

Here we investigate the physical parameters influencing the self-propulsion of a rotating helical filament in a granular medium and discuss the optimal helix design for efficient propulsion. We first consider the propulsive motion of the helix as a function of rotation speed and the influence of the confinement pressure in the medium. The latter is determined via the variation of the depth of the helix within the medium and the applied external load. Subsequently, the problem of a rotating helix is theoretically solved using a frictional model and predictions are compared with experimental data. Finally, considering the results, issues related to the development of a sand-swimming helical are discussed.

\begin{figure}[h!]
\centering
		\includegraphics[width=8.5cm]{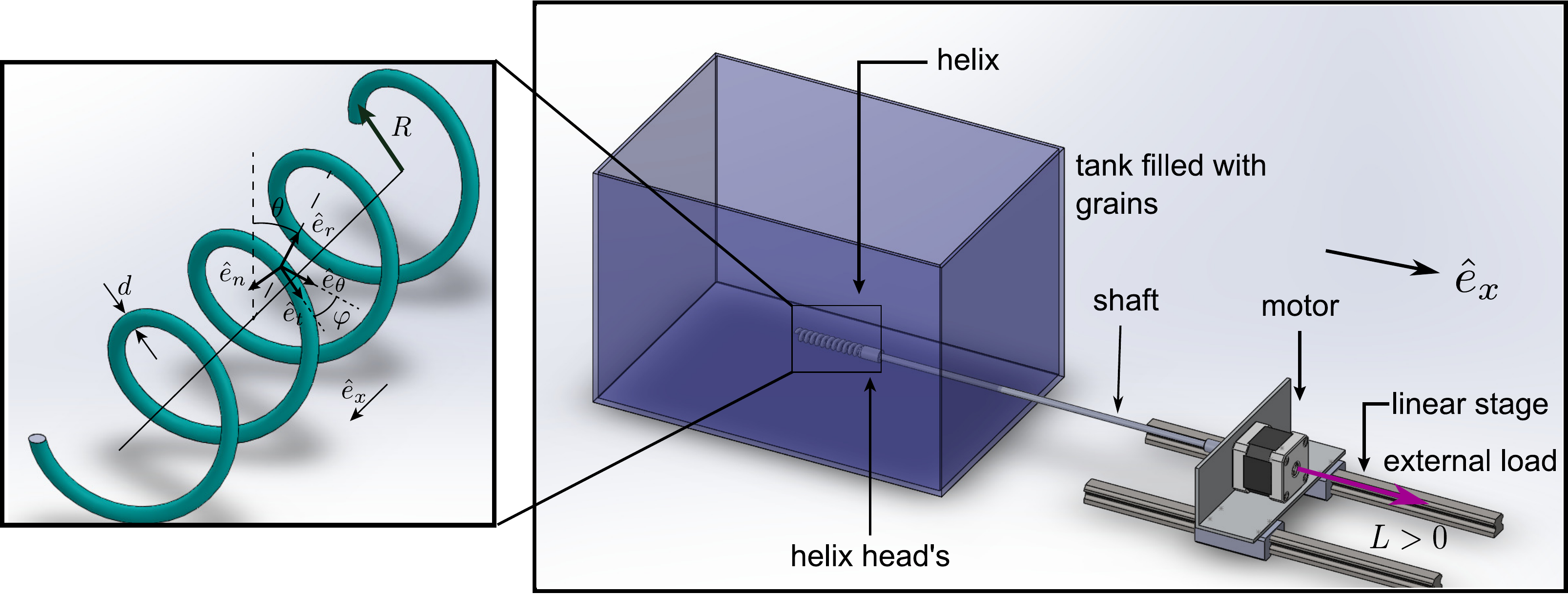}
 	\caption{Sketch of the experimental setup. The left inset presents the helix and vector definitions used in the model.}
		\label{fig:setup}
\end{figure}

The experimental setup consists of a rotating helix placed inside a tank filled with granular materials [Fig. \ref{fig:setup}(a)]. The helix is made of an iron wire of diameter $d=3.0$ mm, has a coil radius of $R= 5.0$ mm, a local inclination $\varphi = 16^\circ$ and a total projected length $\mathcal{L}_x = 58.0$ mm. The tank is a parallel pipe of 30 cm long,  20 cm wide and 20 cm high. The tank is sufficiently large to prevent edge effects and Janssen's effects \cite{seguin2008influence}. The granular medium is composed of plastic beads of diameter $d_g=2.0 \pm 0.1\, \rm{mm}$ and bulk density $\rho_g=2.30 \times 10^3 \, \rm{kg/m^3}$. Efficient compaction of the bed is achieved via gently tapping the tank several times prior to each experiment, a procedure which is recognized to achieve a near random close-packing, $\phi \simeq0.63$ \cite{philippe2002compaction}. Thus, the effective density of the granular material is $\rho = \phi \, \rho_g \approx 1.45 \times 10^3 \, \rm{kg/m^3}$. For effective control of the direction of the propeller motion, the motor providing helix rotation is located on a linear stage, horizontally situated outside the tank. Torque is transmitted to the helix by a thin rod connected to the motor, with the main axis of the helix parallel to the surface. Consequently, the granular pressure remains constant along the trajectory of the propeller \cite{radjai1998features}. Moreover, in order to assess the propeller performance, an external load $L$ is applied to the system composed by the helix, the rod and the motor through a frictionless pulley and a pending mass. The load can be applied either favoring ($L>0$) or opposing ($L<0$) motion. In summary, the experimental configuration allows for the investigation of three parameters, namely, the angular speed of the motor $\omega$, and the height of granular material $h$ overlying the helix and the load $L$. Because the material is opaque, the helix displacement is deduced by tracking the position of the motor outside the tank using a camera, operating at a rate of 25 fps. 

\begin{figure}[h!]
\centering
	\begin{minipage}[c]{0.48\columnwidth}
  		\centering
  		\hspace{0.5cm}(a)\\
		\vspace{0.2cm}
		\includegraphics[width=4.5cm]{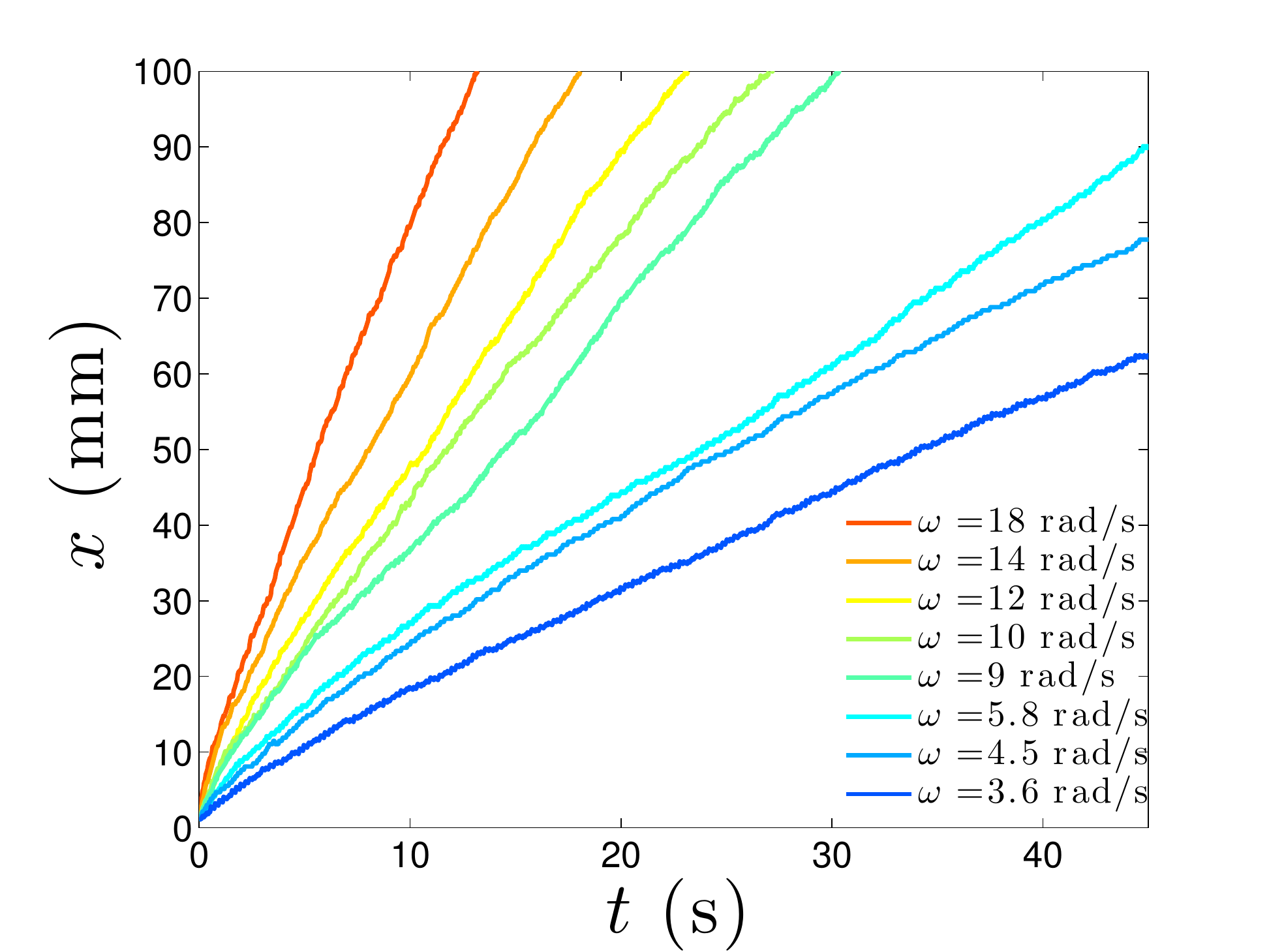}
 	\end{minipage}%
	\begin{minipage}[c]{0.48\columnwidth}
  		\centering
  		\hspace{0.5cm}(b)\\
		\vspace{0.2cm}
		\includegraphics[width=4.5cm]{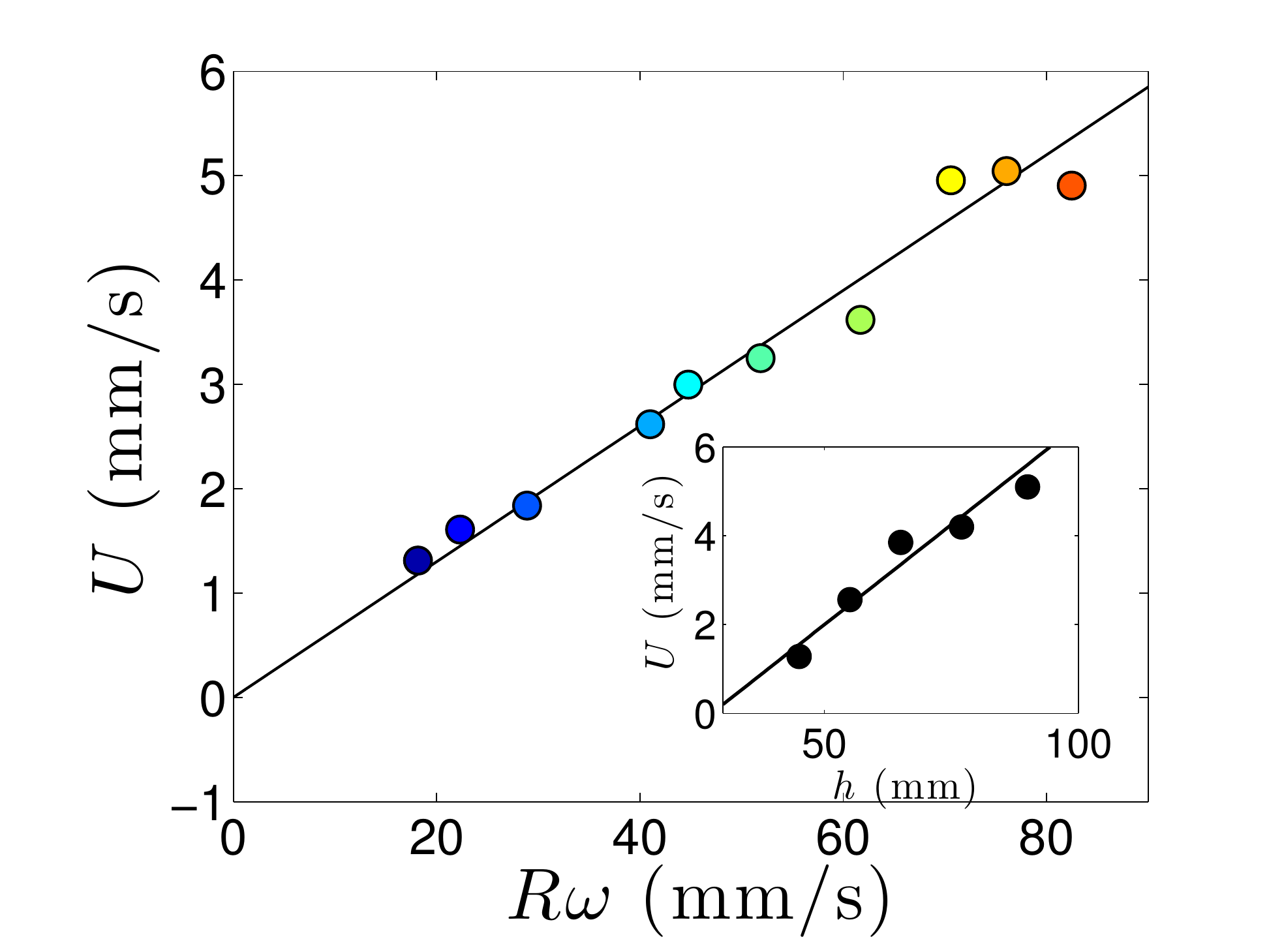}
 	\end{minipage}\\
	\vspace{0.4cm}
	\begin{minipage}[c]{0.48\columnwidth}
  		\centering
  		\hspace{0.5cm}(c)\\
		\vspace{0.1cm}
		\includegraphics[width=4.5cm]{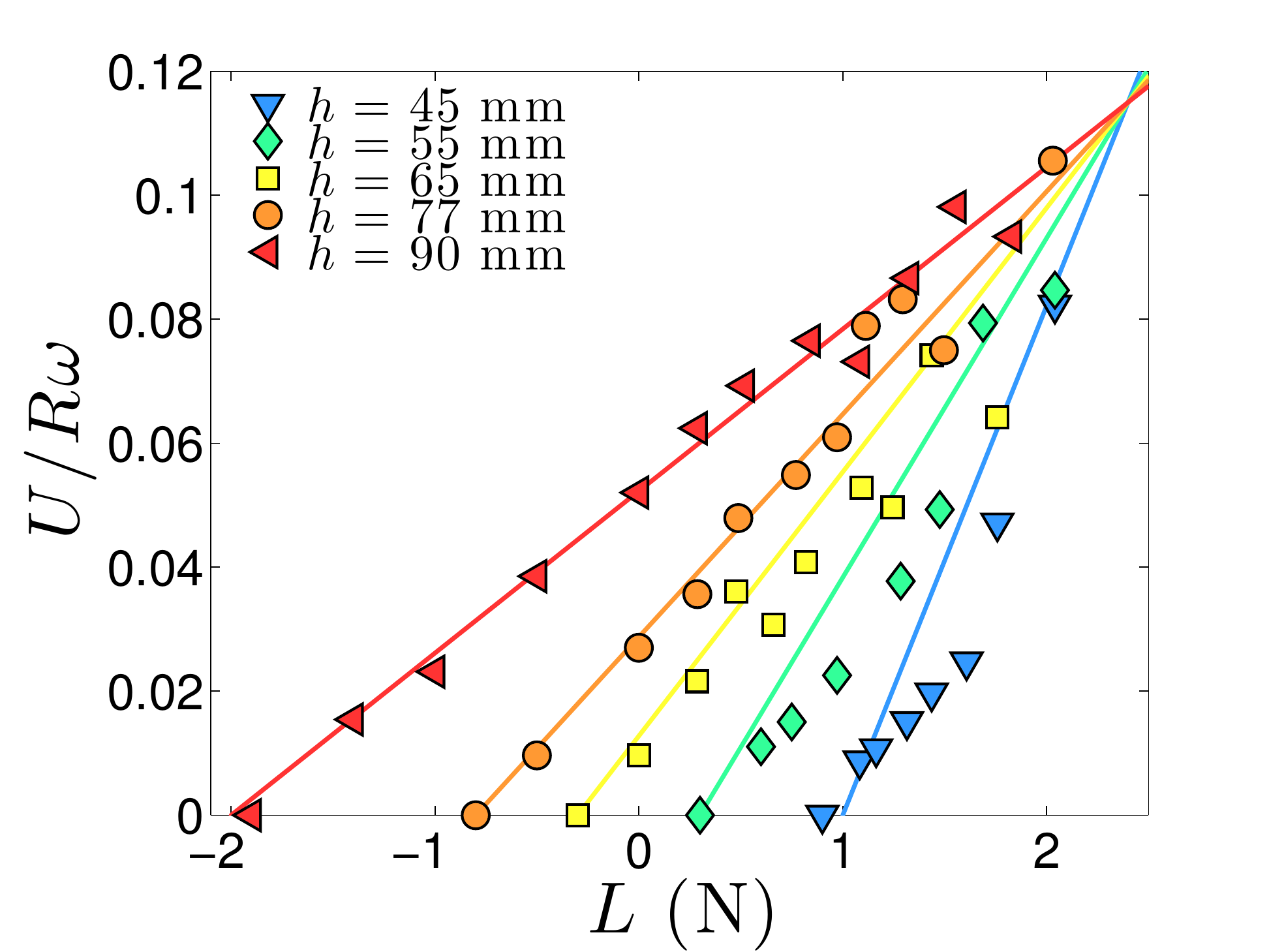}
 	\end{minipage}%
	\begin{minipage}[c]{0.48\columnwidth}
  		\centering
  		\hspace{0.5cm}(d)\\
		\vspace{0.1cm}
		\includegraphics[width=4.5cm]{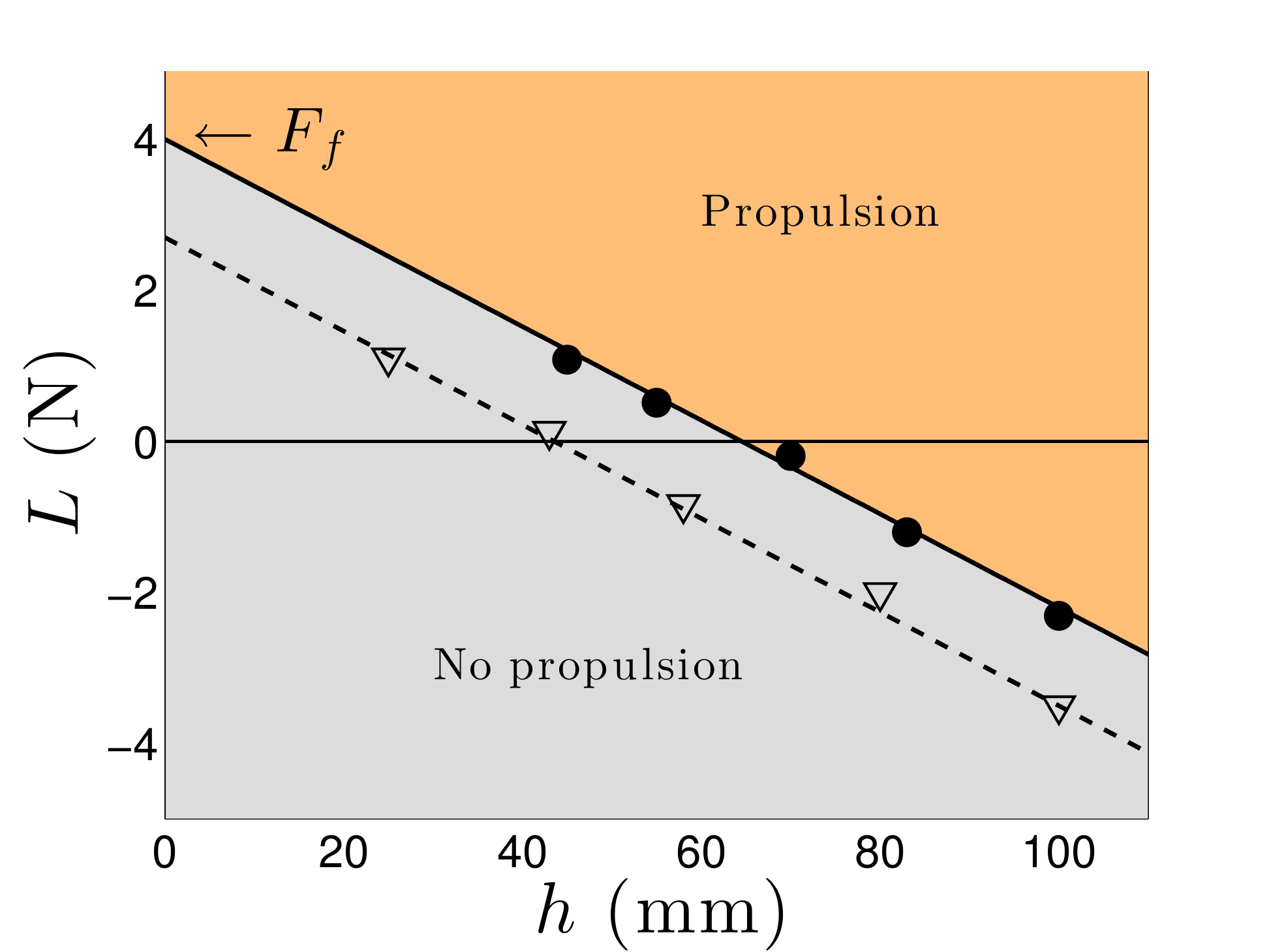}
 	\end{minipage}%
 	\caption{(a) Horizontal position $x(t)$ of the helix as a function of time for different $\omega$, a constant load ($L=1.0$ N) and at a constant depth ($h=75$ mm). (b) $U$ along the $x$-direction as a function  $R \omega$, for a constant load ($L=1.0$ N) and at a constant depth ($h=75$ mm). The inset shows $U$ as a function of $h$, for $L=1.5$ N and $\omega=10.4$ rad/s. (c) Normalized helix velocity $U/R\omega$ as a function of $L$ for a constant $\omega$ ($\omega=10.4$ rad/s). Blue triangles, green diamonds, yellow squares, orange circle and red triangles correspond to experimental data for $h=45$, 55, 65, 77 and 90 mm respectively. Solid lines correspond to Eq. (\ref{eq:helix_speed}) taken from the model results. (d) Black dots show the critical load $L_c$, above which the helix  moves as a function of $h$. Experimental data are interpolated by a linear trend (black solid line). The intersection of this linear fit with the $y$-axis provides an estimate of the internal resistance $F_f$ of the system. This solid line separates the domains of phase space, indicating if helix motion is or is not propulsive. Open triangles indicate the critical load which is measured when friction is reduced by means of air bearings.}
		\label{fig:exp}
\end{figure}

The helix position, $x(t)$, along the horizontal axis, for different rotation speeds and constant load, grows nearly linear with time, 
[Fig. \ref{fig:exp}(a)], indicating that helix motion occurs at constant averaged speed, denoted by $U$. For granular flows, an inertial number can be defined as $I= \dot{\gamma} \, d_g / \sqrt{P /\rho_g }$, where $P$ is the characteristic pressure and $\dot{\gamma}$ the characteristics shear rate \cite{pouliquen2009non}. In the range of parameters under investigation, $I<10^{-3}$, indicating that the flow is quasi-static with negligible inertial effects. 

Subsequently, the effect rotation speed on helix velocity is investigated [Fig. \ref{fig:exp}(b)]. Under constant depth and load, it is apparent that $U$ is proportional to $R \omega$ according to the range of rotation speeds under investigation. This is consistent with the absence of the inertial effect. Furthermore, testing the influence of depth on helix motion shows that $U$ increases linearly with $h$ [with all other parameters kept constant \textit{cf.} inset in Fig. \ref{fig:exp}(b)]. This is related to the increase in the confinement pressure at the propeller, which in turn increases the propelling force.

Finally, when considering the effect of load on the mean speed [Fig. \ref{fig:exp}(c)], it is noted that at low $h$, the helix cannot propel unless assisted by an external load ($L>0$). However, for sufficiently large $h$, the helix propels, even in the presence of an external load opposing its displacement ($L<0$). One observes the presence of a critical load $L_c$, at which the helix becomes stationary. Above $L_c$, the helix propels at a normalized speed $U/R \omega$ that increases with $L$. The value of the critical load $L_c$ is observed to decrease with $h$ [Fig. \ref{fig:exp}(d)]. It is worth noting that the extrapolated value, $F_f$,  of the external load at  $h=0$ (vanishing propulsion) provides the force necessary to overcome both, the friction on the linear stage and the granular drag due to the helix head's. In order to separate these contributions,  experiments with air-lubricated guiding system of negligible friction were performed. The critical load $L_c$ as a function of $h$  (Fig. \ref{fig:exp}(d) open triangles) indicates that using a nearly frictionless system reduces  $F_f$ to 2.5 N, which corresponds solely to the head drag. This drag is known to scale as $\beta \rho g R^2 h$, with $\beta$ a coefficient accounting for both, the properties of the granular medium and the object geometry \cite{albert2001granular}. Thus, the frictional resistance can be reduced through careful design of the form of the objet under propulsion.
 
Moreover, Fig. \ref{fig:exp}(d) defines a phase diagram which draws together all the parameters under which the helical locomotion operates. Finally, the maximal horizontal speed ratio $U/R \omega$ is constant (about 0.11),  [Fig. \ref{fig:exp}(c)], regardless of depth. In the following, it is shown that this maximal speed is solely a function of friction anisotropy and helix geometry.

A simple model is developed in order to rationalize the experimental data. Our model is based on a generalized Coulomb's friction law in which the friction force experienced by a slender cylinder is broken down into components, normal and tangential to the cylinder axis, with two corresponding force coefficients  \cite{maladen2009undulatory, ding2012mechanics}.  
The problem of the helix motion is then described in the cylindrical coordinates system $(\bm{e}_r ,\bm{e}_\theta,\bm{e}_x)$ [Fig. \ref{fig:setup}], and parametrized by $(r=R,\theta, x=R \, \tan \varphi \, \theta)$. A tangential unit vector is introduced $\bm{e}_t = (0, \cos \varphi, \sin \varphi)$ and a normal unit vector $\bm{e}_n= (0, \sin \varphi, - \cos \varphi)$, which define the direct base $(\bm{e}_t,\bm{e}_r,\bm{e}_n)$. The local velocity of a segment of the helix in the cylindrical base is then $\bm{v}=(0, R \, \omega, U)$ and the force per unit length experienced by each segment writes,

\begin{equation}
\bm{f} = - C_t (\bm{e}_v \cdot \bm{e}_t) \bm{e}_t - C_n (\bm{e}_v \cdot \bm{e}_n) \bm{e}_n,
\label{eq:force}
\end{equation}

\noindent where $C_t$ and $C_n$ are the tangential and normal force coefficients, respectively. These coefficients are shown to be proportional to the granular pressure at the object location, the granular packing fraction, and the perimeter (of the cross section) of the slender body \cite{maladen2009undulatory,guillard2013depth}. Maladen \textit{et al.} have developed a refined model based on three  constant coefficients to account for the variation of the friction force with the angle, $\Psi$,  comprised between the cylinder and its velocity. 
Here, we use the simplest frictional law which allows for analytical calculations of the propeller motion, with minimal lost of generality. Indeed, Eq. (\ref{eq:force}) is a first approximation to the formula given by Maladen \cite{maladen2009undulatory} at small  $\Psi$, with the cost of introducing an effective $C_n$ larger than the true value. Thereafter, Eq. (\ref{eq:force}) is used to describe analytically the helix motion under the assumptions of small $\Psi$, small fiber radius ($d \ll R$) and large helix step ($R \tan \varphi \gg d_g$). From Eq. (\ref{eq:force}), each segment of the helix experiences a force that reads,

\begin{equation}
\bm{f} = - \frac{ C_t (R \, \omega \cos \varphi + U \sin \varphi ) \bm{e}_t + C_n ( R \, \omega \sin \varphi - U \cos \varphi ) \bm{e}_n  }{ \sqrt{U^2 + ( R \omega )^2}}.
\label{eq:force2}
\end{equation}

\noindent Equation (\ref{eq:force2}) implies that the sum of local forces along the helix path only has a component in the $x$-direction, $F_x$, which writes, $F_x = \int_0 ^\mathcal{L} \bm{f} \cdot \bm{e}_x  \, ds$, $\mathcal{L}$ being the total length of the helical filament. 

In a steady regime, the propulsive force $F_x$ is balanced by the drag force, $F_d$.  $F_d$ includes all the experimental device losses, $F_f$,  and the external load, $L$;  $F_d=F_f - L$. Thus, the helix speed verifies,

\begin{widetext}
\begin{equation}
 \frac{F_d}{\mathcal{L}_x (C_n - C_t) \cos \varphi} \sqrt{1+ \left( \frac{R\omega}{U} \right)^2} - \frac{R \omega}{U} + \frac{C_t \tan \varphi + C_n / \tan \varphi}{C_n - C_t}=0,
 \label{eq:eq_speed}
\end{equation}
\end{widetext}

\noindent where $\mathcal{L}_x=\mathcal{L} \sin \varphi$ is the projected helix length along the $x$-direction. Introducing non-dimensional parameters $\tilde{U}= U/R\, \omega$, $\tilde{F}=F_d/\mathcal{L}_x (C_n - C_t) \cos \varphi$ and $\tilde{U}_m=(C_n - C_t)/(C_t \tan \varphi + C_n / \tan \varphi)$, Eq. (\ref{eq:eq_speed}) writes,

\begin{equation}
(1- \tilde{F}^2) \left( \frac{1}{\tilde{U}} \right)^2 - \frac{2}{\tilde{U}_m \tilde{U}} +  \left( \frac{1}{\tilde{U}_m} \right)^2 - \tilde{F}^2= 0,
\label{eq:helix_speed_eq} 
\end{equation}

\noindent which accepts an analytical solution in the form,

\begin{equation}
\tilde{U}= \frac{\tilde{U}_m (1-\tilde{F}^2)}{1 + \sqrt{1+ (1-\tilde{F}^2) (\tilde{F}^2 \tilde{U}_m ^2 -1)}}.
\label{eq:helix_speed}
\end{equation}

Thus, the helix speed is proportional to the rotation speed, a fact which is consistent with experiential observations [Fig. \ref{fig:exp}(b)]. 
It can be observed in Eq. (\ref{eq:helix_speed}) that $\tilde{U}\rightarrow 0$ for $\tilde{F}\rightarrow1$ and $\tilde{U}\rightarrow \tilde{U}_m$ for $\tilde{F}\rightarrow 0$. Demonstrating that without resistance ($F_d =0$), the helix reaches the maximal speed $U_m= \tilde{U}_m R \omega$ whereas, when the resistance equals the maximal propulsive force ($F_m=\mathcal{L}_x (C_n - C_t) \cos \varphi$), the helix's speed decreases to zero. Equation (\ref{eq:helix_speed}) also predicts that the maximal propulsive force is proportional to the projected helix length and to the friction force anisotropy. In addition, the model provides the expression of the maximal speed of the helix, as a function of the friction force coefficients $C_t$, $C_n$ and the helix angle $\varphi$. Interestingly, the expression of the maximal speed implies that $ \tilde{U}_m$ tends to $\tan \varphi$ if the ratio $C_n/C_t$ tends to an infinite value; a situation where the helix follows its own path. 

Finally, theoretical predictions are compared with experimental data. The solution of Eq. (\ref{eq:helix_speed}) is reported in Fig. \ref{fig:exp}(c), indicated by solid lines with the parameter values $\tilde{F}$ and $\tilde{U}_m$ determined experimentally. It is observed that the experimental data agree with the model and it is noted that the solution of Eq. (\ref{eq:helix_speed}) shown in Fig. \ref{fig:exp}(c) is similar to a linear trend. Indeed, in the case where $\tilde{U}_m \ll 1$, Eq. (\ref{eq:helix_speed}) can be elaborated upon at the first order to provide $\tilde{U}= \tilde{U}_m (1 - \tilde{F})$. Furthermore, the linear fit of the experimental data in Fig. \ref{fig:exp}(d) means that the following is established as $C_n-C_t = c \, d \, \rho g  h$ with $c\simeq 24$. This value can be compared with the one determined by Guillard \textit{et al.} for a cylinder stirred into a granular medium, written as $c\simeq 15$, and thus consistent with our measurements \cite{guillard2013depth}.

\begin{figure}[h!]
\centering
	\begin{minipage}[c]{0.7\columnwidth}
  		\centering
  		\hspace{0.5cm}(a)\\
		\vspace{0.1cm}
		\includegraphics[width=5.2cm]{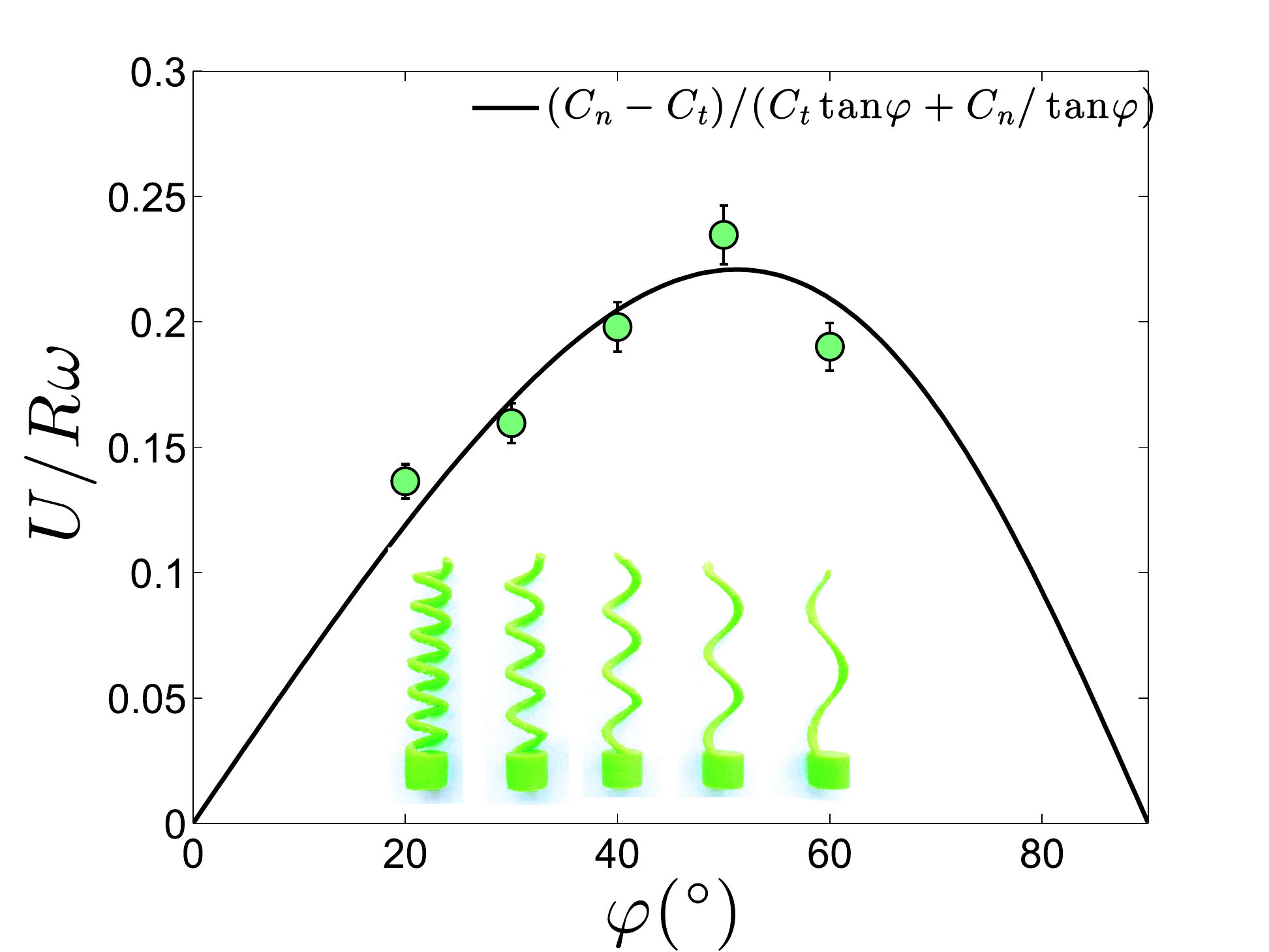}
 	\end{minipage}%
	\begin{minipage}[c]{0.28\columnwidth}
  		\centering
  		\hspace{0.2cm}(b)\\
		\vspace{0.2cm}
		\includegraphics[width=1.8cm]{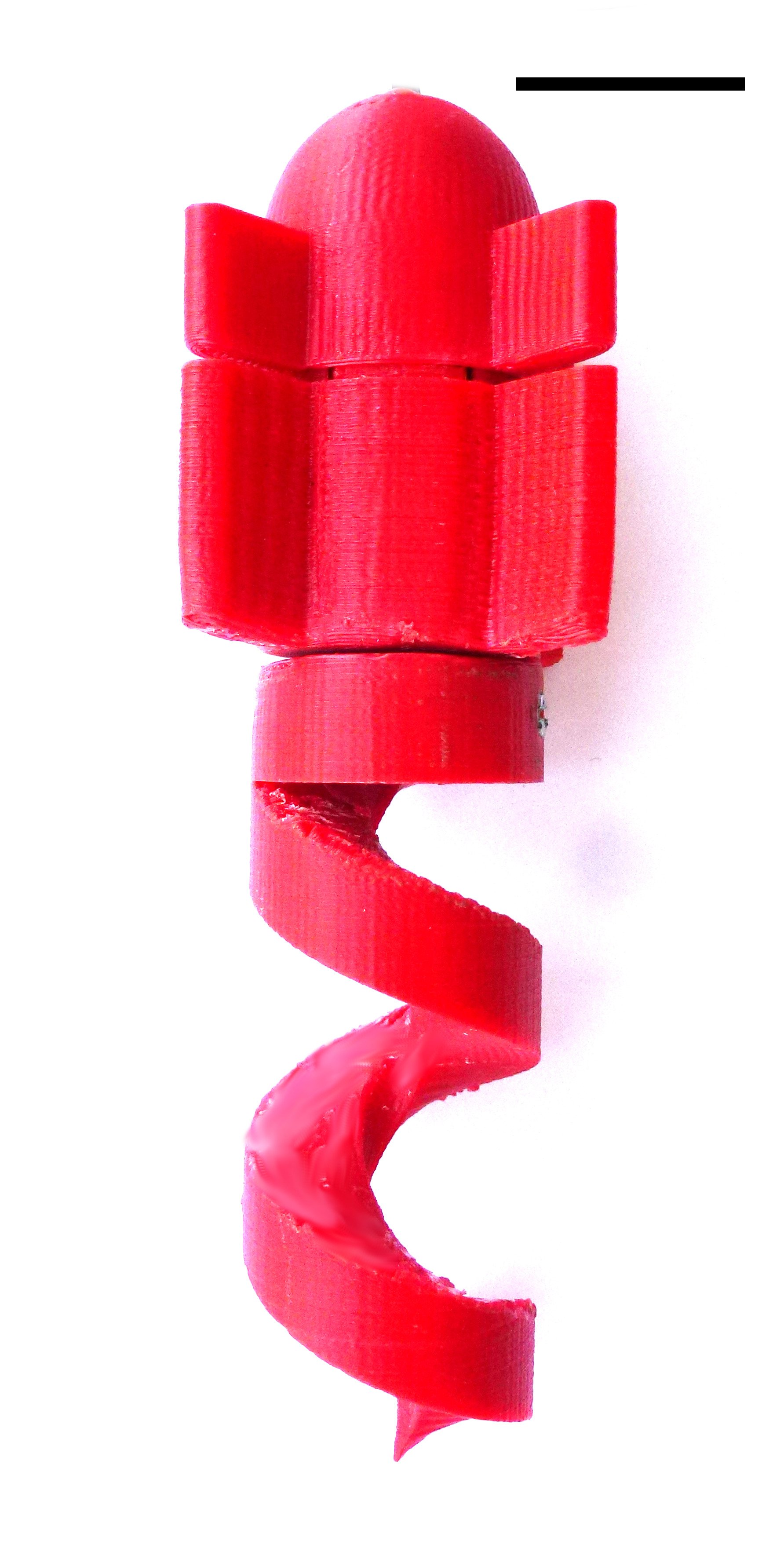}
		\vspace{0.2cm}
 	\end{minipage}
 	\caption{(a) $U/R \omega$ vs $\varphi$ for $\mathcal{L}_x=$ 58 mm, $R=5.0$ mm,  $d=3.0$ mm and $h=50$ mm.  To compensate for head's drag, a load $L=1.2$N is applied. The solid line shows the best fit to the experimental data with the relation $\tilde{U}_m=(C_n - C_t)/(C_t \tan \varphi + C_n / \tan \varphi)$ and $C_n/C_t=1.6$.  Inset: 3D printed helices of varying $\varphi$. (b) Helical robot for motion in a granular medium. Black bar is 20 mm.}
		\label{fig:helix_angle}
\end{figure}

According to the theoretical model, the maximal normalized velocity of the helix in the bed, reached at the vanishing load, is $\tilde{U}_m=(C_n - C_t)/(C_t \tan \varphi + C_n / \tan \varphi)$. This expression indicates the existence of a maximal value of the normalized speed relative to the helix angle $\varphi$. It is observed that in order to maximize the speed, the optimal value of the helix angle $\varphi_o$ is verified as $\tan \varphi_o=\sqrt{C_n/C_t}$. For $C_n/C_t=2$, the following is obtained $\varphi_o = 54.7^\circ$. To check this prediction, we 3D-printed five helices with distinct angles $\varphi$. A good agreement with the maximum speed law is observed with a friction coefficient ratio $C_n/C_t=1.6$ [Fig. \ref{fig:helix_angle}(a)]. Due to the fact that the optimal helix angle for rapid propulsion is only dependent on the force coefficient ratio, it is implied that the locomotion depth, the packing fraction and the radius of the helix wire do not influence the optimum speed. However, the maximal propulsive force is an increasing function of these variables.

The experimental values of the maximal propulsive force $F_m$ and its maximal normalized speed $\tilde{U}_m$, produce the coefficients $C_t$ and $C_n$ within this system. Indeed, the ratio of these coefficients is expressed as,

\begin{equation}
\frac{C_n}{C_t} = \frac{1+ \tilde{U}_m \tan \varphi}{1 - \tilde{U}_m/ \tan \varphi}
\label{eq:friction_coef_ratio}
\end{equation}

The helix with a local slope of $\varphi=16^\circ$, was observed experimentally to have $\tilde{U}_m= 0.11$. According Eq. (\ref{eq:friction_coef_ratio}), these values lead to $C_n/C_t \simeq 1.7$, which are compatible with measurements from Malden \textit{et al.} \cite{maladen2009undulatory}. $C_t$ and $C_n$ are sensitive to granular preparation,  doubling their values from loose to close packing \cite{maladen2009undulatory}. Our model thus accounts for different granular preparations, with the condition of adjusting these coefficients in accordance with the material packing.

The total power required to move the helix is defined as, $\mathcal{P} = \int_0 ^\mathcal{L} \bm{f} \cdot \bm{v} \, ds$. The fraction of this power that contributes to propulsion is, $\mathcal{P}_x = \int_0 ^\mathcal{L} \bm{f}_x \cdot \bm{v} \, ds$. As a consequence, the efficiency of the process writes, $\eta=\mathcal{P}_x/\mathcal{P}$ and can be expressed analytically as a function of $\tilde{F}$, $C_n/C_t$ and $\varphi$. For $C_n/C_t=1.7$ and $\varphi=16^\circ$, it was found that $\eta$ reaches a maximal value of about 4\% for $\tilde{F}=0$. In the case where $\tilde{F}=0$ and $C_n/C_t=1.7$, $\eta$ reaches a maximal value of 40 \% for $\varphi=59^\circ$. Otherwise, the power efficiency is observed to increase with the friction coefficient ratio $C_n/C_t$.

These findings on helical propulsion could be used in the development of a sand-robot. For effective propulsion, the robot must abide by two conditions. First, the maximal propulsive force ($F_m$) must be larger than the resistive one ($F_d$). The latter is dominated by the granular drag on the robot's head and can be reduced by optimization of the head's shape, head rotation \cite{jung2017reduction} and vibration \cite{texier2017low}.
Secondly, the torque required to rotate the helix has to be lower than the one required to rotate the head of the helix, otherwise the helix will stand still. 

To verify this theory, a robot respecting these constraints has been designed and operated in granular media, as shown in Fig. \ref{fig:helix_angle}(b). It includes a hemispherical head containing a battery (Lithium 3.7 V 50 mAh) and a motor (DC 4 V with a gear reduction 298:1). A helical tail of length 45 mm and local inclination $\varphi=30^\circ$ is linked to the axis of the motor. Four pallets prevent head rotation but ensure helix rotation. All the plastic components of the robot were printed with a domestic 3D printer. This robot has been tested in various granular media (sand, glass and plastic beads) with successful propulsion in all media. The normalized mean speed of the robots under a constant granular pressure is observed to remain constant regardless the grain size, consistently with the frictional approach.

This study affirms that the rotating helix's forward velocity is dependent on the rotation speed, resistive force and helix geometry.  Based on these findings, it was possible to obtain the optimal helical shape for propulsion in a non-cohesive medium. A proof-of-concept robot showed effective propulsion in a variety of granular materials. In conclusion, the present study proves that locomotion in granular media   is dictated by the anisotropy of friction and the symmetry breaking of the helical shape. Moreover, despite physical differences between granular material and viscous fluid, locomotion in both media appear to share the same physical origin, suggesting that the generalization of Purcell's scallop theorem to granular media would be possible.

\begin{acknowledgments}
B.D.T.  acknowledges the support of the CONICYT-Chile postdoctoral fellow Fondecyt: 3160167.  F. M. is grateful to Anillo ACT-1412 for additional support. The authors sincerely thank Manouk Abkarian for beneficial discussions, Tania Sauma for careful reading and Nelson Flores for valuable help in the realization of the experimental setup. 
\end{acknowledgments}

\end{document}